\documentclass[12pt,a4paper,english]{article}
\usepackage{babel,cite} \usepackage[dvips]{epsfig}

\makeatletter

\usepackage{amssymb,amsthm}

\newcommand{\be}{\begin{equation}} 
\newcommand{\ee}{\end{equation}}
\newcommand{\eq}[1]{(\ref{#1})}

\def\nn{\nonumber} 
\def\bea{\begin{eqnarray}} 
\def\eea{\end{eqnarray}}
\newcommand{\barr}{\begin{array}} 
\newcommand{\earr}{\end{array}}
 
\def\one{\mbox{1 \kern-.59em {\rm l}}}

\def\({\left(} 
\def\){\right)} 
\def\[{\left[} 
\def\]{\right]}

\def\a{\alpha}  
  
 \def\d{\delta}

  \def\la{\lambda}  
 \def\O{\Omega}

\def\R{{\mathbb R}}  
 \def\one{\mbox{1 \kern-.59em {\rm l}}}

\def\cH{{\cal H}}

\def\bit{\begin{itemize}} \def\eit{\end{itemize}}

\def\diag{\mbox{diag}}

   \def\dd{\partial}

\makeatother
\begin{document}

\renewcommand{\title}[1]{\vspace{10mm}\noindent{\Large{\bf
#1}}\vspace{8mm}} \newcommand{\authors}[1]{\noindent{\large
#1}\vspace{5mm}} \newcommand{\address}[1]{{\itshape #1\vspace{2mm}}}

\begin{flushright}
UWThPh-2005-30
\end{flushright}

\begin{center}

\title{ \Large Renormalization of the noncommutative $\phi^3$ model \\[1ex]
through the Kontsevich model}

\vskip 3mm

\authors{Harald {\sc Grosse} and 
Harold {\sc Steinacker\footnote{Supported by the FWF project P16779-N02.}}}

\vskip 3mm

\address{Institut f\"ur Theoretische Physik, Universit\"at Wien\\
Boltzmanngasse 5, A-1090 Wien, Austria}

\vskip 1.4cm

\textbf{Abstract}

\vskip 3mm

\begin{minipage}{14cm}%

We point out that the noncommutative selfdual $\phi^3$ model
can be mapped to the Kontsevich model, 
for a suitable choice of the eigenvalues in the latter. 
This allows to apply
known results for the Kontsevich model 
to the quantization of the 
field theory, in particular the KdV flows and Virasoro
constraints.
The 2-dimensional case is worked out explicitly.
We obtain nonperturbative expressions for the genus expansion of
the free energy and some $n$-point functions. 
The full renormalization for finite coupling
is found, which is determined by the genus 0 sector only. 
All contributions in a genus expansion of any $n$-point function
are finite after renormalization.
A critical coupling is determined beyond which the model
is unstable. The model is free of UV/IR diseases.

\end{minipage}

\end{center}

\section{Introduction}

One of the motivation for
noncommutative field theory is the hope
to achieve a better understanding
of the UV divergences and renormalization in 
quantum field theory (QFT), by formulating QFT on a quantized 
or noncommutative (NC) space. The most popular example of such a 
space is the quantum plane, where the coordinate 
``functions'' satisfy the canonical commutation relations
$ [x_i,x_j] = i \theta_{ij}$. This introduces 
a length scale, and divides space essentially into 
``Planck cells'' of finite area. At first sight, one might then guess
that this length scale corresponds to a cutoff 
$\Lambda_{NC} = \frac 1{\sqrt{\theta}}$ in QFT.
However, it turns out that $\Lambda_{NC}$ does not 
play the role of a UV cutoff,
but more properly serves as a reflection point between scales in the
UV and the IR on both sides of $\Lambda_{NC}$. Moreover,
this generically 
leads to the so-called UV/IR  mixing in divergent QFT's, 
which is a serious obstacle to perturbative renormalization 
\cite{Minwalla:1999px}.

A way to overcome these problems has been found recently in 
\cite{Grosse:2003nw,Grosse:2004yu,Rivasseau:2005bh} 
for the scalar $\phi^4$ model 
in 2 and 4 dimensions. This is achieved by
adding a confining potential (a ``box'') to the field theoretical 
models, i.e. an additional relevant term in the Lagrangian
of the form $\Omega (\tilde x_i \phi)(\tilde x_i \phi)$ which makes them
covariant under a duality \cite{Langmann:2002cc}.
This additional term essentially
suppresses or controls the divergencies in the IR, and introduces 
discreteness into the models which 
can be viewed as generalized matrix models.
Perturbative renormalizability was then proved 
using a renormalization group approach.
In particular, there is a special point $\Omega =1$ where the models
become selfdual in a certain sense \cite{Langmann:2002cc}. 
This is expected to be preserved under renormalization, 
and is therefore of particular interest.

The close relationship between noncommutative field theory and 
(generalized) matrix models can be seen in many ways. 
The most striking similarity is that  
Feynman diagrams are drawn on a Riemann surface, leading to
a genus expansion. 
For generic momenta, only planar diagrams are divergent just 
like in ordinary matrix models. However, for exceptional momenta
 the nonplanar (higher-genus) diagrams become divergent as well,
which leads to UV/IR mixing i.e. new divergences in the IR.
This similarity to matrix models suggests to apply or adapt
some of the powerful techniques which have been developed
in the context of matrix models. This idea was applied 
in \cite{Langmann:2003if}, where the selfdual complex $\phi^4$
model was formulated as matrix model, which in the degenerate
$U(N)$-invariant case was solved exactly but turned out to be trivial
(i.e. free).
Later, in \cite{Steinacker:2005wj,Steinacker:2005tf} 
a strategy to analyze more general
scalar models using  matrix model techniques was proposed.
This is based on the hypothesis that the eigenvalue distribution
is localized as in ordinary matrix models, and
strongly hints at nontriviality of e.g. the real $\phi^4$
model. 

In the present paper, we show that the 
noncommutative Euclidean selfdual $\phi^3$ model
can be solved (at least in 2 dimensions, but probably more generally) 
using matrix model techniques, 
and is related to the KdV hierarchy. 
This is achieved by rewriting 
the field theory as Kontsevich matrix model, for a suitable choice
of the eigenvalues in the latter. The relation holds 
for any even dimension, and allows to apply some of the known, remarkable 
results for the Kontsevich model to the quantization of the 
$\phi^3$ model. We work out the 2-dimensional case explicity. 
This allows to write down closed 
expressions for the genus expansion of the free energy, 
and also for some $n$-point functions by taking derivatives and
using the equations of motion. 
It turns out that the required renormalization 
is determined by the genus 0 sector only, 
and can be computed explicitly. We show that
using this renormalization, 
all contributions in a genus expansion of any $n$-point function
correlation function are finite and well-defined for finite coupling.  
This implies but is stronger than perturbative renormalization.
Here we draw heavily on results of 
\cite{Itzykson:1992ya,Kontsevich:1992} for the Kontsevich model.  

We thus obtain
fully renormalized models with nontrivial interaction
which are free of UV/IR diseases. 
Only the linear tadpole term in the action 
requires (logarithmic) renormalization, while  mass and coupling
constant do not run as expected. We also extract the 
leading perturbative contribution for certain
correlators using the nonperturbative results, 
and verify that they coincide with the 
diagrammatic computations.
All this shows that even though the $\phi^3$ may appear 
ill-defined at first, it is in fact much better under control
than other models.

These results are obtained starting with purely imaginary coupling 
constants, but allow analytic continuation 
to real coupling. This turns out to be well-defined for 
finite coupling below some critical value, but becomes unstable for 
strong coupling. We derive the corresponding critical coupling,
which is interpreted as
instability induced by the finite potential barrier.

In view of the well-known relation between NC field theory and string theory
in certain backgrounds \cite{Seiberg:1999vs},
our results are relevant also in that context.
In fact, the noncommmutative $\phi^3$ model has been related to 
open strings in a $B$-field background in \cite{Andreev:2000rm}. 
More recently, the relevance of the Konsevich model 
to open string field theory \cite{Gaiotto:2003yb}
and to branes on non-compact
Calabi-Yau spaces \cite{Aganagic:2003qj} has been pointed out. This
suggests that the language and techniques of NC field theory
might be useful also in this string-theoretical context.

This paper is organized as follows: 
In section \ref{sec:phi3} we define the $\phi^3$ model
under consideration, and rewrite it as 
Kontsevich model. Some useful identities for correlators are 
also given. We then recall the most important facts about the
Kontsevich model in section \ref{sec:kontsevich}, and specialize it to the 
field theoretical model under consideration in section \ref{sec:appplic}.
Renormalization and  finiteness are established in section
\ref{sec:renormaliz}, which is the main result of this paper. 
We then perform some perturbative checks, and 
provide some approximation formulas valid for finite coupling. 
The critical point for real coupling is studied in section
\ref{sec:critical}, and some perturbative computations are given in section 
\ref{sec:perturbative}. We conclude with a discussion and outlook.

\section{The noncommutative $\phi^3$ model}
\label{sec:phi3}

We assume some familiarity with the formulation of noncommutative 
field theory, as reviewed e.g. in \cite{Douglas:2001ba,Szabo:2001kg}.
Consider the action 
\be 
\tilde S = \int_{\R^{2n}_{\theta}} \frac 12 \partial_i\phi
\partial_i\phi + \frac {\mu^2}2 \phi^2 + \Omega^2 (\tilde x_i \phi)
(\tilde x_i \phi) + \frac{i\tilde\lambda}{3!}\;\phi^3 
\ee 
on the $2n$-dimensional quantum plane, which is generated by self-adjoint
operators\footnote{One can of course also start with 
the Moyal product on a classical space of functions. 
We will skip this point of view here, since for our purpose
the matrix representation is crucial.} $x_i$
satisfying the canonical commutation relations
\be 
[x_i,x_j] = i \theta_{ij}.  
\label{CCR}
\ee 
We also introduce 
\be 
\tilde x_i = \theta^{-1}_{ij} x_j, \qquad [\tilde x_i,\tilde
x_j] = i \theta^{-1}_{ji}
\ee 
assuming that $\theta_{ij}$ is 
nondegenerate.
The dynamical object is the scalar field
$\phi = \phi^\dagger$, which is  a self-adjoint operator acting
on the representation space $\cH$ of the algebra \eq{CCR}.
The term $\Omega^2 (\tilde x_i \phi)(\tilde x_i \phi) $ is included following 
\cite{Grosse:2003nw,Langmann:2002cc}, making the model covariant under 
Langmann-Szabo duality, and taking care of the UV/IR mixing.
We choose to write the action with an imaginary coupling $i\tilde \la$, 
assuming $\tilde \la$ to be real. The reason is that for real coupling
$\tilde \la' = i\tilde \la$, the potential 
would be unbounded from above and below, 
and the quantization would seem ill-defined. We will see 
however that 
the quantization is completely well-defined for imaginary 
$i\tilde \la$, 
and allows an analytic continuation to real $\tilde \la' = i\tilde \la$
in a certain sense which will be made precise below.
Therefore we accept for now that the action $\tilde S$ is not
real.

Using the commutation relations \eq{CCR}, the derivatives $\partial_i$
can be written as inner derivatives
$\partial_i f = -i[\tilde x_i,f]$.
Therefore the action can be
written as 
\bea 
\tilde S &=& \int -\frac 12 [\tilde x_i,\phi][\tilde
x_i,\phi] + \Omega^2 \tilde x_i \phi \tilde x_i \phi + \frac {\mu^2}2
\phi^2 + \frac{i\tilde \lambda}{3!}\;\phi^3 \nn\\ 
&=& \int -(\tilde
x_i\phi\tilde x_i\phi - \tilde x_i \tilde x_i \phi\phi) + \Omega^2
\tilde x_i \phi \tilde x_i \phi + \frac {\mu^2}2 \phi^2 +
\frac{i\tilde \lambda}{3!}\;\phi^3 
\eea 
using the cyclic property of the
integral. 
For the ``self-dual'' point $\O =1$, this action simplifies
further to 
\be 
\tilde S = \int (\tilde x_i \tilde x_i) \phi\phi + \frac
{\mu^2}2 \phi^2 + \frac{i\tilde \lambda}{3!}\;\phi^3.
\label{action-E} 
\ee 
In order to quantize the theory,  we need to 
introduce a
counterterm which is linear in $\phi$: 
\be 
\tilde S = \int (\tilde
x_i \tilde x_i) \phi\phi - (i\tilde \la \tilde a) \phi 
+ \frac {\mu^2}2 \phi^2 +
\frac{i\tilde \lambda}{3!}\;\phi^3.
\label{action-E-full} 
\ee 
We will include this additional term from now on, which is written as 
$(i\tilde \la \tilde a)$ since then only the coupling 
$\tilde \la' = (i\tilde \la)$ 
needs analytic continuation, while $\tilde a$ will always be real.
The additional term implies that the potential $V(\phi) =
\tilde \la' \tilde a \phi + \frac {\mu^2}2 \phi^2 
+ \frac{\tilde \la'}{3!}\;\phi^3$ has a local minimum at 
$\phi_0 = \frac 1{\tilde\la'} \big(-\mu^2+ \sqrt{\mu^2-2(\tilde \la')^2
  \tilde a}\big)$ for real $\tilde \la'$. 

The crucial observation which we will use in this paper is the 
fact that the self-dual model \eq{action-E-full} 
can be written as a matrix model 
coupled to an external ``source'' matrix. 
This was already noted and applied in \cite{Langmann:2003if} 
in the context of the complex $\phi^4$ model. 
To see this, consider the operator 
\be 
J = 4\pi (\theta\, \sum_i \tilde x_i \tilde x_i + \frac
{\mu^2\theta}2 ).
\ee 
We  assume that $\theta_{ij}$ has the canonical form 
$\theta_{12} = -\theta_{21} =: \theta \,\,(= \theta_{34} = -\theta_{43}$ etc. for higher dimensions),
focusing on the 2-dimensional case in this paper. 
Then $J$ is essentially 
the Hamiltonian of quantum mechanical harmonic oscillator
$J \propto \sum \hat x^2 + \hat p^2+ const $, 
which in the usual basis of
eigenstates diagonalizes with evenly spaced eigenvalues, 
\be 
J|n\rangle = 4\pi(n\;+ \frac {1 + \mu^2\theta}2)|n\rangle, \qquad n \in
\{0,1,2,...  \};
\label{J-explicit}
\ee 
we will only consider the 2-dimensional case from now on.
Replacing $\int = (2\pi \theta) Tr$, the action
can  be written as 
\be 
\tilde S = 2\pi Tr\Big( \frac 1{4\pi} J \phi^2 +
\frac{i\tilde \lambda\theta}{3!}\;\phi^3 
- i\tilde \la \tilde a \theta\phi\Big).  
\label{action-tilde} 
\ee 
It is
convenient to define the dimensionless constants
\be 
\la = 2\pi \theta\tilde \lambda, \, \qquad
a = \la^2 \tilde a = -(i\la)^2 \tilde a,
\label{const-defs}
\ee 
and to subtract an irrelevant constant
term $ Tr(\frac 1{3\la^2} J^3 + \frac 1{\la^2} J a)$ from the action,
in order to simplify some formulas below. Then we
obtain the following action
\be 
S := \,Tr \Big( \frac 12 J
\phi^2 + \frac{i \la}{3!}\;\phi^3 - \frac{a}{i\la} \phi - \frac 1{3(i\la)^2}
J^3 - \frac {1}{(i\la)^2} J a\Big).
\label{action-kontsevich}
\ee 
One can eliminate the quadratic term by shifting\footnote{for the
  quantization, the integral
for the diagonal elements is then defined via analytical continuation,
and the off-diagonal elements remain hermitian since $J$ is diagonal.}
the  variable in \eq{action-kontsevich} as 
\be 
\tilde\phi =
\phi + \frac 1{i\la} J,
\ee 
so that
\be 
S = Tr \Big( -\frac 12 (
\frac 1{i\la} J^2 +\frac{2a}{i\la}) \tilde\phi + \frac{i
\la}{3!}\;\tilde\phi^3 \Big) 
= Tr \Big( -\frac 1{2 i \la} M^2 \tilde\phi + \frac{i
\la}{3!}\;\tilde\phi^3 \Big)
\label{action-kontsevich-lin}
\ee 
where 
\be 
M = \sqrt{J^2 + 2a}.
\label{M-def}
\ee 
Now the field $\tilde\phi$
couples linearly rather than quadratically to the source, which
turns out to be very useful.  Alternatively, the linear term can
be eliminated by setting 
\be 
\tilde \phi = X + \frac 1{i\la} M, \qquad X
= \tilde \phi - \frac 1{i\la} M = \phi + \frac{J-M}{i\la}.
\ee 
Then the action becomes
\be 
S = \,Tr \Big(\frac 12 M X^2 + \frac{i\la}{3!}\;X^3 -
\frac 1{3(i\la)^2} M^3 \Big),
\label{action-kontsevich-N}
\ee 
which has the form of the Kontsevich model \cite{Kontsevich:1992}.

\subsection{Quantization and equations of motion}
\label{sec:quantization}

The quantization of the model \eq{action-E-full} 
resp. \eq{action-kontsevich-lin} is defined by an integral over
all Hermitian $N\times N$ matrices $\phi$, where $N$ serves as a UV
cutoff. 
The partition function is defined as
\be 
Z(M) = \int D\tilde \phi \exp(- S(M))
\ee 
for $S = S(M)$ given by \eq{action-kontsevich-lin}, and correlators 
or ``$n$-point functions'' are defined through
\be
\langle \phi_{i_1 j_1} ....  \phi_{i_n j_n}\rangle
= \frac 1Z\, \int D\tilde \phi \exp(- S)\,
\phi_{i_1 j_1} ....  \phi_{i_n j_n}
\label{correl-def}
\ee
The eigenvalues of $J$ resp. $M$ are then given by \eq{J-explicit} 
resp. \eq{M-def}
for $n=0,1,..,N-1$.
The nontrivial task is to show that all correlation functions have 
a well-defined 
and hopefully nontrivial limit $N \to \infty$, i.e. that the
``low-energy physics'' is well-defined and
independent of the cutoff.

Using 
the symmetry $Z(M) = Z(U^{-1} M U)$ for $U \in U(N)$, 
we can assume that 
$M$ is diagonalized with (ordered) eigenvalues $m_i$. 
Then there is a residual $U(1)^{N}$ invariance 
$\phi_{ij} \to u_i^{-1} \phi_{ij} u_j$ with $u_i \in U(1)$.
This implies certain obvious ``index conservation laws'',
e.g. $\langle\phi_{kl} \rangle = \delta_{kl} \langle\phi_{ll} \rangle$
etc.

In order to have a well-defined limit $N\to\infty$, 
we should require in particular that the 2-point function 
$\langle\phi_{ij}\phi_{kl}\rangle $ and 
also the one-point function $\langle\phi_{kl} \rangle$ 
have a well-defined limit.
We therefore impose the renormalization conditions 
\bea 
&&\langle\phi_{00} \phi_{00}\rangle = \frac 1{2\pi} \frac 1{\mu^2\theta +1}, 
\label{renorm-cond-0}\\ &&
\langle\phi_{00} \rangle =0 
\label{renorm-cond}
\eea 
which hold in the free case $\la =0$.  
This will uniquely determine the renormalization of $a$ 
(and $\mu^2$, which receives only finite quantum corrections and
which will not be computed here).

\paragraph{Quantum equations of motion and correlators.}
 
We first derive some simple relations for the basic correlators.
Using the identity\footnote{this is justified by analytic continuation 
using \eq{action-kontsevich-N}.}
\be 
0= \int
d\tilde \phi {d \over d\tilde \phi_{kk}} \exp Tr (\frac 1{2i\la} \tilde
\phi M^2 - \frac{i\la}{3!} \tilde \phi^3) 
\ee 
we obtain
\be 
0=
\Big<\sum_{l,\, l\ne k}\tilde \phi_{kl}\,\tilde \phi_{lk} +
\tilde\phi_{kk}^2 -\frac 1{\la^2} m_k^2 \Big>
\label{Sdyson-1}
\ee 
for each matrix index $k$.
Insertions of a diagonal factor $\tilde
\phi_{kk}$ in the correlators
can be achieved\footnote{Alternatively one could also
promote $a$ to a matrix (commuting with $J$), and take derivatives
w.r.t. $a_i$. Since this presents no particular advantages, we will not
do this in the following.}
 by acting with
the derivative operator $ {i\la\over m_k}\frac{\dd{}}{\dd m_k}$ 
on $Z$. More general non-diagonal insertions 
$\tilde \phi_{kl}$ will be discussed in section 
\ref{sec:general-correl}.

In the Kontsevich model resp. the
$\phi^3$ model, there is a simple way to obtain also a  certain class of
non-diagonal insertions, 
by expressing the invariance of the integral $Z$
under an infinitesimal change of variable of the form 
\cite{Itzykson:1992ya}
\be
\tilde \phi \to \tilde \phi+i\varepsilon [X_{(kl)},\tilde \phi], {\rm \quad with \quad }
(X_{(kl)})_{ab}=\delta_{ak}\delta_{bl}\tilde \phi_{kl}
\ee
which gives explicitly 
\be
\delta_{(kl)} \tilde \phi_{ij} = i \varepsilon 
\left(\d_{ik} \tilde \phi_{il} \tilde \phi_{lj} 
- \d_{jl} \tilde \phi_{ik} \tilde \phi_{kj}\right).
\ee
The Jacobian is $1+i\varepsilon (\tilde \phi_{ll}-\tilde \phi_{kk})$,
while the term $Tr \tilde \phi^3$ is invariant. Thus
\be
0= \Big< \tilde \phi_{ll}-\tilde \phi_{kk}+{1\over 2i\la} (m_k^2-m_l^2)\tilde \phi_{kl}\tilde \phi_{lk}
\Big>,
\label{Sdyson-2}
\ee
with no summation implied.
In particular, we find for the propagator
\be
\Big<\tilde \phi_{kl}\tilde \phi_{lk}\Big> = 
 \frac {2i\la}{m_k^2-m_l^2} \Big<\tilde\phi_{kk}-\tilde \phi_{ll}  \Big>
\label{Sdyson-3}
\ee
for $k \neq l$ (no sum). 
Recalling that $\tilde\phi = \phi + \frac 1{i\la}J$, this gives
\bea
\Big<\phi_{kl}\phi_{lk}\Big> &=& 
 \frac{2i\la}{m_k^2-m_l^2} (\Big<\phi_{kk}-\phi_{ll}  \Big> + \frac 1{i\la}(J_k - J_l)) \nn\\
&=& \frac{2}{J_k + J_l} + \frac{2i\la}{m_k^2-m_l^2} \Big<\phi_{kk}-\phi_{ll}  \Big> 
\label{2point-eom}
\eea
noting that $J_k^2 - J_l^2 = m_k^2 - m_l^2$.
The first term is the free contribution, and the second the
quantum correction.
Thus we ``only'' need the 1-point functions
\be
\Big<\tilde \phi_{kk}\Big> 
= {i\la\over m_k}\frac{\dd{}}{\dd m_k } \,\ln \tilde Z(m)  
= \frac 1{i\la} J_k  + \Big<\phi_{kk}\Big>.
\label{linear-expect}
\ee
They can be obtained from the Kontsevich model, as we will show 
in detail.

Proceeding as in \cite{Itzykson:1992ya}, 
we can insert \eq{Sdyson-2} into \eq{Sdyson-1} which leads to
\be
\frac{m_k^2}{\la^2} = -\langle \tilde \phi_{kk}^2 \rangle 
 -(2i\la) \sum_{l, l\ne k}
{\langle \tilde \phi_{kk}- \tilde \phi_{ll}\rangle \over
  m_k^2-m_l^2}.
\label{phikk2}
\ee
These manipulations can be generalized: 
consider
\bea
0 &=& \d_{(kl)} Z \left<\tilde \phi_{kk}\right> 
= \d_{(kl)} \int D\tilde \phi \exp (-S(J)) \tilde \phi_{kk} \nn\\
&=& Z\left<\left((\tilde \phi_{ll} - \tilde \phi_{kk}) +  {1\over 2i\la}
(m_k^2-m_l^2)\tilde \phi_{kl}\tilde \phi_{lk}\right) \tilde \phi_{kk}
 + \d_{(kl)}  \tilde \phi_{kk}\right> \nn\\
&=&Z\left<\left((\tilde \phi_{ll} - \tilde \phi_{kk}) +  {1\over 2i\la}
(m_k^2-m_l^2)\tilde \phi_{kl}\tilde \phi_{lk}\right) \tilde \phi_{kk}
 + \left(\tilde \phi_{kl} \tilde \phi_{lk} 
- \d_{kl} \tilde \phi_{kk} \tilde \phi_{kk}\right)\right> 
\eea
using
\be
\delta_{(kl)} \tilde \phi_{kk} = i \varepsilon 
\left(\tilde \phi_{kl} \tilde \phi_{lk} 
- \d_{kl} \tilde \phi_{kk} \tilde \phi_{kk}\right).
\ee
This implies
\bea
\left<\tilde\phi_{kl}\tilde\phi_{lk} \tilde \phi_{kk}\right>
&=& \frac{2i\la}{m_k^2-m_l^2}\, 
\left<(\tilde \phi_{kk} - \tilde \phi_{ll})\tilde \phi_{kk}
 - \left(\tilde \phi_{kl} \tilde \phi_{lk} 
- \d_{kl} \tilde \phi_{kk} \tilde \phi_{kk}\right)\right> \nn\\
&=&  \frac {2i\la}{m_k^2-m_l^2}\, 
\left<(1+\d_{kl}) \tilde \phi_{kk}\tilde \phi_{kk}
 - \tilde \phi_{ll}\tilde \phi_{kk}
- \frac {2i\la}{m_k^2-m_l^2}\,(\tilde \phi_{kk} - \tilde \phi_{ll}) \right> \nn\\
\label{3point-1}
\eea
(no sum). 
Therefore 
$\left<\tilde \phi_{kl}\tilde \phi_{lk}\tilde \phi_{kk} \right>$
is finite and known exactly
provided the 1- and 2-point functions are known, which 
indeed will be obtained from the Kontsevich model. 

Clearly these manipulations can be generalized. 
However, we will present a different argument in section 
\ref{sec:general-correl}
which establishes finiteness of general correlation functions 
more directly, using the renormalization procedure explained below.
The relation \eq{3point-1} could
also be used to demonstrate that the model is not free,
in contrast to \cite{Langmann:2003if}. 
We will not bother to elaborate this, since 
the lowest nontrivial term in a Taylor expansion in $\la$ 
is manifestly finite 
and nonzero, and the model will be renormalized for finite coupling.

It is worth pointing out that
\eq{phikk2} yields the matrix Airy equations  
\be
\Big[ m_k^2 -(\frac 1{m_k}\frac{\dd{}}{\dd m_k})^2
-2 \sum_{l, l\ne k}
{1\over m_k^2-m_l^2}(\frac 1{m_k}\frac{\dd{}}{\dd m_k}-\frac
1{m_l}\frac{\dd{}}{\dd m_l})
\Big]Z=0
\label{mastereq}
\ee
as shown in \cite{Itzykson:1992ya},
which was called ``Master equation'' in \cite{Makeenko:1991ec}.
This in turn can be translated into the Virasoro constraints,
\be
L_m Z =0, \qquad m \geq -1
\label{virasoro}
\ee
for suitable operators $L_m$.
It is also useful to define as in \cite{Makeenko:1991ec,Langmann:2003if} 
\be
W(m_k) := \frac{1}{m_k} \frac{\partial{}}{\partial m_k } \,\ln  Z(m)
 = \frac{1}{i\la} \langle \tilde \phi_{ii}\rangle.
\label{W-def}
\ee
Then \eq{mastereq} can be
written as
\be
W(m_k)^2 + \frac{1}{m_k} \frac{\dd{}}{\dd m_k } \,W(m_k)
+2 \sum_{l \ne k} \frac{W(m_k)- W(m_l)}{m_k^2-m_l^2} - m_k^2 = 0.
\label{master-2}
\ee
Before proceeding, we briefly recall the usual treatment \cite{Makeenko:1991ec}
of the genus 0 solution of the Kontsevich model. In that case one 
can neglect the term 
$\frac{1}{m_k} \frac{\dd{}}{\dd m_k } \,W(m_k)$  in \eq{master-2},
which would be suppressed by $\frac 1N$. Then 
\eq{master-2} constitutes $N$ equations in $N$ unknowns 
$W(m_k),\, k=1,..., N$. This can be solved 
in the large $N$ limit, by assuming that
$W(m_k)$ becomes an analytic function with cuts in the large $N$ limit,
where $m_k,\, k=1,..., N$ becomes a continuous variable
(recall that $m_k$ are the ordered eigenvalues of $M$).

In the present case, 
we are not allowed to neglect the term 
$\frac{1}{m_k} \frac{\dd{}}{\dd m_k } \,W(m_k)$ 
since there is {\em no} factor $N$ involved in \eq{W-def}. 
Correspondingly 
$\langle \tilde \phi_{kk}^2 \rangle \neq \langle \tilde \phi_{kk}\rangle^2$,
as can be seen explicitly in perturbation theory
(see section \ref{sec:perturbative}).
Therefore we have to use the complete genus expansion, which is indeed
available more-or-less explicitly. It will turn out that only the
genus 0 contribution requires renormalization.
We start by recalling some more facts about the Kontsevich model.

\section{Some useful facts for the Kontsevich model}
\label{sec:kontsevich}

The Kontsevich model is defined by
\be
 Z^{Kont}(\tilde M)=e^{F^{Kont}}=
\frac{\int dX \exp\left \{Tr \left(-\frac{\tilde M X^2}{2}
+i \frac{X^3}{6} \right)\right\} }
{\int dX \exp\left\{-Tr \left(\frac{\tilde M X^2}{2}\right)\right\} }
\label{Z-Konts}
\ee
where $\tilde M$ is a given hermitian $N \times N$ matrix,
and the integral is over Hermitian $N \times N$ matrices $X$.
This model has been introduced by Kontsevich \cite{Kontsevich:1992} 
as a combinatorial way of computing certain topological
quantities (intersection numbers) on moduli spaces of Riemann surfaces
with punctures, which in turn were related to the partition function
of the general one-matrix model by Witten \cite{Witten:1990hr}.
It turns out to have an extremely rich
structure related to integrable models (KdV flows) and
Virasoro constraints,
and was studied using a variety of techiques. For our purpose, the most 
important results are those of 
\cite{Kontsevich:1992,Makeenko:1991ec,Itzykson:1992ya} which provide
explicit expressions for the genus expansion of 
the free energy of \eq{Z-Konts}.
Note that $\la$ can be introduced via
\bea
Z^{Kont}(\tilde M) &=&
Z^{Kont}(\la^{-2/3}  M) =
\frac{\int dX \exp\left \{Tr \left(-\la^{-2/3} \frac{M X^2}{2}
+i\frac{X^3}{6} \right)\right\} }
{\int d X \exp\left\{-Tr \left(\la^{-2/3} \frac{ M X^2}{2}\right)\right\}}\nn\\
&=&\frac{\int d\tilde X \exp\left \{Tr \left(-\frac{\ M \tilde X^2}{2}
-i\la\frac{\tilde X^3}{6} \right)\right\} }
{\int d\tilde X \exp\left\{-Tr \left(\frac{M \tilde X^2}{2}\right)\right\}
},
\label{Z-Konts-2}
\eea
where $X = -\la^{1/3}  \tilde X,\, M = \la^{-2/3}\tilde  M$, 
which allows to obtain the analytic
continuation in $\la$.

The matrix integral in \eq{Z-Konts} and its large $N$ limit
can be defined rigorously 
in terms of its asymptotic series.
Defining the normalized measure
\be
d\mu_{\tilde M}(X) ={dX \exp( -\frac 12 Tr \tilde M X^2) \over
\int dX \exp (-\frac 12 Tr \tilde M X^2) } 
\ee
for Hermitian $N\times N$ matrices,
one considers the matrix Airy function
\bea
Z^{Kont \,(N)} &=& \int d\mu_{\tilde M}(X) \exp (\frac i 6 Tr X^3) 
=\sum_{k\ge 0}
Z^{Kont\,(N)}_k(\tilde M) \nn\\
Z^{Kont \,(N)}_k(\tilde M) &=&{(-1)^{k}\over (2k)!} \int d\mu_{\tilde M}(X)
\left({Tr X^3\over 6}\right)^{2k}.
\label{Zk-def}
\eea
A crucial fact \cite{Kontsevich:1992} 
is that the terms $Z_k^{Kont\,(N)} (\tilde M) $ can be 
expressed as {\em polynomials}
with rational coefficients in the variables
\be
\theta_r= {1\over r} Tr \tilde M^{-r}.
\ee
Then $Z^{Kont\,(N)}_k$ is homogeneous of degree $3k$, if we set
\be
\deg \theta_r = r.
\label{deg-def}
\ee
Only the first $N$ of the $\theta_r$ are algebraically independent
for an $N\times N$ matrix $\tilde M$. However, it is known 
\cite{Itzykson:1992ya,Kontsevich:1992} that 

{\em Considered as a function of $\theta_{\textstyle .}
\equiv\{\theta_1,\theta_2,...\}$,
$Z_k^{Kont\,(N)} (\tilde M)$ is independent of $N$ for $3k \le N$ 
and depends only on
$\theta_r$, $1\le r\le 3k$.} 

This allows one to define unambiguously the series
$Z^{Kont}(\theta_i) = \sum_{k\ge 0} Z^{Kont}_k(\theta_i)$
where
$Z^{Kont}_k(\theta_i) := Z^{Kont\,(N)}_k (\theta_i), \,\, N \ge 3k$
without any further reference to $N$. 
Even though we formally work with an infinite number of 
variables $\theta_r$, 
 each $Z^{Kont}_k$ ($Z_0=1$) depends on finitely
many of them.

Furthermore, the following remarkable facts hold 
\cite{Kontsevich:1992}:
\begin{enumerate}
\item
\be
\frac{\partial Z^{Kont}}{\partial \theta_{2r}} =0 
\ee
\item
$Z^{Kont}(\theta_r)$ is a $\tau$-function for the Korteweg-de Vries
equation. 
\end{enumerate}
Namely if
\bea
t_r &=& -(2r+1)!!\,\,\theta_{2r+1}= -(2r-1)!!\,\,tr \tilde M^{-2r-1}\nn\\
u &:=&  \frac{\partial^2}{\partial t_0^2 } \ln Z^{Kont}
\label{var-defs}
\eea
then
\be
\frac{\partial  u}{\partial t_1} = \frac{\partial}{\partial t_0}\bigg( {1\over 12} {\partial^2 u\over
\partial t_0^2 }+\frac 12 u^2 \bigg)
\ee
and more generally
\be
{\partial\over \partial t_n}u ={\partial \over\partial t_0}R_{n+1}.
\ee
Here the $R_n$ denote the Gelfand-Dikii differential polynomials
(derivatives are taken with respect to $t_0$)
\bea
R_2&=& {u^2\over 2}+{u''\over 12} \nn\\
R_3&=& {u^3\over 6}+{uu''\over 12}+{u'^2\over 24}+{u^{(4)}\over 240}\nn\\
R_4&=& {u^4\over 24}+{uu'^2\over 24}+{u^2u''\over 24}+{uu^{(4)}\over 240}
+{u'u'''\over 120}+{(u'')^2\over 160}+{u^{(6)}\over 6720}\nn\\
& \ldots  \nn\\
R_n&=& {u^n\over n!}+\cdots \nn\\
\eea
computed from
\be
(2n+1)R'_{n+1}=\frac 1{4}R'''_n+2u R'_n +u' R_n.
\ee
For the present case, we have
\be
\theta_r 
= {\la^{2r/3}\over r} \sum_{n\geq 0} \frac 1{(J_n^2 \, +2a)^{r/2}}.
\ee
Without renormalization (i.e. for finite or zero $a$), 
$\theta_r$ is logarithmically divergent for $r=1$, and finite 
for $r \geq 2$. This is a first indication that the model
requires renormalization, and we will see that $a \propto \la^2 \ln N$. 
However, it will turn out that even in the properly renormalized case
a different set of variables is more suitable.

A further crucial fact is the existence of a  
\paragraph{Genus expansion.}

As usual for matrix models,
one can consider the genus expansion 
\be
\ln Z^{Kont} = F^{Kont} = \sum_{g\geq 0} F^{Kont}_g
\ee
by drawing the Feynman diagrams on a suitable Riemann surface.
In principle, 
this genus expansion can be obtained
as a $\frac 1N$ expansion by introducing
an explicit factor $N$ in the action, 
so that the action takes the form
\be 
S' = \,Tr N ( -\frac 12 M'^2 \phi' +
\frac{1}{3!}\; \phi'^3).
\label{action-kontevich-Nprime}
\ee 
However, it was shown in \cite{Itzykson:1992ya} that
the $F^{Kont}_g$ can also be
computed  using the KdV equations and the Virasoro constraints \eq{virasoro}, 
which allows to find closed expressions
for small $g$. 
It is useful to use the following set of variables:
\be
I_k(u_0,t_i) = \sum_{p \geq 0} t_{k+p} \frac{u_0^p}{p!}
\label{I-k-1}
\ee
where $u_0$ is given by the solution of the implicit equation
\be
u_0 = I_0(u_0,t_i).
\ee
We note that using the definition \eq{var-defs}, $I_k$ can be resummed as
\be
I_k(u_0,t_i) = - (2k-1)!! \sum_{i\geq 0} \frac 1{(\tilde m_i^2 - 2
  u_0)^{k+\frac 12}},
\label{I-k-2}
\ee 
in particular
\be
u_0 = -\sum_{i\geq 0} \frac{1}{\sqrt{\tilde m_i^2-2u_0}} = I_0.
\label{u-constraint}
\ee
These variables turn out to be more useful for our purpose than the 
$t_r$, since the $\tilde m_i^2 - 2 u_0$ will be finite in the
renormalized model, while the $t_r$ are not.
Using the KdV equations, 
\cite{Itzykson:1992ya} found the following explicit formulas:
\bea 
F^{Kont}_0 &=& {u_0^3\over 6}-\sum_{k\ge 0}{u_0^{k+2}\over k+2}{t_k\over k!}
+\frac 12 \sum_{k\ge 0}{u_0^{k+1}\over k+1}
    \sum_{a+b=k}{t_a\over a!}{t_b\over b!}  \label{F-0-IZ}\\
F^{Kont}_1 &=& \frac 1{24} \ln \frac 1{1-I_1},  \label{F-0-IZ-1}\\\
F^{Kont}_2 &=& \frac 1{5760}\left[5{I_4\over (1-I_1)^3} +29{I_3 I_2\over
(1-I_1)^4 } +28{I_2^3\over (1-I_1)^5}\right]\ , 
\label{F-higher}
\eea
etc. All $F^{Kont}_g$ with $g \geq 2$ are  given by
{\em finite} sums of polynomials in $I_k/ (1-I_1)^{{2k+1\over 3}}$, the
number of which is $p(3g-3)$ with $p(n)$ being the number of partitions of
$n$.
Expanded in the original variables $t_r$, this becomes \cite{Itzykson:1992ya}
\bea
F^{Kont}_0 
&=& \frac{t_0^3}{3!} + t_1 \frac{t_0^3}{3!}  + \(t_2 \frac{t_0^4}{4!}
+ 2 \frac{t_1^2}{2!} \frac{t_0^3}{3!}\)
 + \(t_3 \frac{t_0^5}{5!} + 3 t_1 t_2 \frac{t_0^4}{4!} + 6 \frac{t_0^3}{3!}
\frac{t_1^3}{3!}\)\nn\\
&& + \[t_4 \frac{t_0^6}{6!} +\(6 \frac{t_2^2}{2!} + 4 t_1 t_3\)
 \frac{t_0^5}{5!} + 24 \frac{t_0^3}{3!} \frac{t_1^4}{4!} +
12 t_2 \frac{t_1^2}{2!} \frac{t_0^4}{4!}\]  \nn\\
&&  +\[t_5 \frac{t_0^7}{7!}+\(5 t_1 t_4 + 10 t_2 t_3\)\frac{t_0^6}{6!}
+ 120 \frac{t_0^3}{3!} \frac{t_1^5}{5!} +\(30 t_1 \frac{t_2^2}{2!}
 + 20 t_3 \frac{t_1^2}{2!}\)\frac{t_0^5}{5!} + 60 t_2 \frac{t_1^3}{3!}
\frac{t_0^4}{4!}\]   \nn\\
&& + \ldots 
\label{F0-explicit}
\eea
and
\bea
24F^{Kont}_1 
 &=& t_1 +\(\frac{t_1^2}{2!} + t_0 t_2\)
 +\(2 \frac{t_1^3}{3!} + t_3 \frac{t_0^2}{2!} + 2 t_0 t_1 t_2\) \nn\\
&&  +\(6 \frac{t_1^4}{4!} + t_4 \frac{t_0^3}{3!} + 4 \frac{t_0^2}{2!}
\frac{t_2^2}{2!} + 6 t_0 t_2 \frac{t_1^2}{2!} + 3 t_1 t_3
\frac{t_0^2}{2!}\)\nn\\
&&  +\(24 \frac{t_1^5}{5!} + t_5 \frac{t_0^4}{4!} + 24 t_0 t_2
\frac{t_1^3}{3!}+\(4 t_1 t_4 + 7 t_2 t_3\)\frac{t_0^3}{3!}
+ 16 t_1 \frac{t_0^2}{2!} \frac{t_2^2}{2!}
+ 12 t_3 \frac{t_0^2}{2!} \frac{t_1^2}{2!}\)  \nn\\
&& + \ldots 
\label{F1-explicit}
\eea
An alternative, more useful form of $F^{Kont}_0$
can be obtained by solving directly the ``master-equation'' 
\eq{master-2} at genus 0, 
as discussed in section \ref{sec:quantization}. 
This leads to \cite{Makeenko:1991ec}
\bea
F_0^{Kont}&=&\frac{1}{3}\sum_{i=0}^{N} \tilde m_i^3 -
\frac{1}{3}\sum_{i=0}^{N} (\tilde m_i^2-2u_0)^{3/2}
-u_0 \sum_{i=0}^{N}(\tilde m_i^2-2u_0)^{1/2} \nonumber \\
&&+ \frac{u_0^3}{6}-\frac{1}{2}
\sum_{i,k=0}^{N}\ln\left\{\frac{(\tilde m_i^2-2u_0)^{1/2}+(\tilde m_k^2-2u_0)^{1/2}}
{\tilde m_i+\tilde m_k}\right\}
\label{F0Kont}
\eea
which is equivalent to \eq{F-0-IZ} but more useful in our context
because it gives explicitly the analytic continuation. 
The parameter $u_0$ is again given by the implicit equation \eq{u-constraint}.
This constraint can alternatively
be obtained by considering  $F^{Kont}(\tilde m_i;u_0)$
with $u_0$ as an independent variable, since its equation of motion
\be
\frac{\partial}{\partial u_0}F^{Kont}(\tilde m_i;u_0) 
= \frac 12 (u_0 - I_0)^2 =0
\label{constr-implicit}
\ee
reproduces the constraint. 
All sums now range from $0$ to $N$, and will be convergent 
after renormalization as $N\to\infty$
 for the physical observables.

\section{Applying Kontsevich to the $\phi^3$ model}
\label{sec:appplic}

We need 
\be
Z = Z^{Kont}[\tilde M] Z^{free}[\tilde M]\exp(\frac 1{3(i\la)^2} Tr M^3)
\ee
where
\be
Z^{free}[\tilde M]= e^{F_{free}}=
\int dX \exp\left(-Tr \left(\frac{\tilde M X^2}{2}\right)\right) 
= \prod_i \frac 1{\sqrt{\tilde m_i}}\,\prod_{i<j} \frac 2{\tilde m_i+\tilde m_j}
\ee
up to  irrelevant constants, so that
\be
F_{free} = - \frac 12\sum_{i, j=1}^N 
\ln(\tilde m_i+\tilde m_j) \quad (+ const).
\label{F-free}
\ee
Therefore
\bea
F_0 &:=& F_0^{Kont} + F_{free} + \frac 1{3(i\la)^2} Tr M^3 \nn\\
&=&  - \frac{1}{3}\sum_{i=0}^{N} \sqrt{\tilde m_i^2-2u_0}^3
-u_0 \sum_{i=0}^{N}\sqrt{\tilde m_i^2-2u_0} \nonumber \\
&&+ \frac{u_0^3}{6}-\frac{1}{2}
\sum_{i,k=0}^{N}\ln(\sqrt{\tilde m_i^2-2u_0}+\sqrt{\tilde m_k^2-2u_0}).
\label{F0}
\eea
In the present case, the eigenvalues $\tilde m_i$ are given by 
\eq{Z-Konts-2}, \eq{M-def}
\be
\tilde m_i = \la^{-2/3}\sqrt{J_i^2 +2a},
\ee
and the model will be ill-defined without renormalization 
since $u_0$ is
logarithmically divergent. However, we note that only 
the combinations $\sqrt{\tilde m_i^2 -2u_0}$ enter in \eq{I-k-2} and \eq{F0}, 
which can be rewritten as 
\be
\sqrt{\tilde m_i^2 -2u_0} 
= \la^{-2/3}\sqrt{J_i^2 + 2(a - \la^{4/3}u_0)}  
= \la^{-2/3}\sqrt{J_i^2 + 2b}
\ee
where
\be
b =  a - \la^{4/3}u_0 = a - \la^{4/3}I_0.
\label{b-def}
\ee
The point is that $b$ will be finite after renormalization,
which makes the model well-defined.

\paragraph{Analytic continuation and elimination of $u_0$.}

Replacing $u_0$  by \eq{b-def}, 
the genus 0 contribution to the partition function \eq{F0} 
takes the form
\bea
F_0 &=& \ln Z_{g=0} 
=
-\frac{\la^{-2}}{3}\sum_{i=0}^{N} \sqrt{J_i^2+2b}^3
-\la^{-2}(a-b)\sum_{i=0}^{N}\sqrt{J_i^2 +2b} \nonumber \\
&&+ \frac{\la^{-4}}{6}(a-b)^3 
-\frac{1}{2}
\sum_{i,k=0}^{N}\ln\left(\la^{-2/3}\sqrt{J_i^2+2b}
                        +\la^{-2/3}\sqrt{J_k^2+2b}\right).
\label{F0tilde-2}
\eea
We consider $F = F(J)$ as a function of 
(the eigenvalues of) $J$ from now on.
$b$ satisfies
the implicit constraint \eq{b-def}, which becomes
\be
(b -a) =  \la^2\sum_{i=0}^N \frac 1{\sqrt{J_i^2 + 2 b}}\,\, .
\label{constraint-3}
\ee
In particular, the $\tilde m_k$ can be analytically continued
as long as $\sqrt{J_i^2+2b}$ is well-defined.
Imposing the constraint explicitly we have
\bea
F_0 &=& -\frac{\la^{-2}}{3}\sum_{i=0}^{N} \sqrt{J_i^2+2b}^3
-\left(-\sum_{k=0}^N \frac{1}{\sqrt{J_i^2+2b}}\right) 
\sum_{i=0}^{N}\sqrt{J_i^2 +2b} \nonumber \\
&&+ \frac{1}{6}\la^2
\left(-\sum_{k=0}^N \frac{1}{\sqrt{J_i^2+2b}}\right)^3 
-\frac{1}{2}
\sum_{i,k=0}^{N}\ln\left(\la^{-2/3}\sqrt{J_i^2+2b}
                        +\la^{-2/3}\sqrt{J_k^2+2b}\right).\nn\\
\label{F0tilde-3}
\eea
Now the model depends only on the $J_k$, and $b$
will be fixed together with the counterterm $a$
 by the renormalization condition \eq{renorm-cond} and the constraint 
\eq{constraint-3}.
Also, note that the unexpected $\la^{-2}$ dependence is 
spurious due to the term 
 $-\frac{\la^{-2}}{3} Tr J^3$ in \eq{action-kontsevich}, which has no physical
 significance an could be subtracted.

For some computations it is useful to consider $b$ 
in \eq{F0tilde-2} as an independent
auxiliary variable, which then satisfies the constraint 
through the e.o.m as in \eq{constr-implicit}.
The essential observation is
\be
\frac{\partial}{\partial  J_i}  F_0(J_i)
 = \frac{\partial}{\partial J_i} F_0(J_i;b)
 +  \frac{\partial}{\partial b} F_0(J_i;b)
  \frac{\partial}{\partial J_i} b 
=\frac{\partial}{\partial J_i} F_0(J_i;b) 
\label{Fkont-partial}
\ee
using
\be
\frac{\partial}{\partial b} \, F_0(J_i;b) 
= -\frac 12 \Big(\la^{-2}(b-a)  - \sum_{i=0}^N\frac 1{\sqrt{J_i^2 + 2 b}}\,\Big)^2 =0,
\label{constr-implicit-2}
\ee
due to the constraint  \eq{constraint-3}.

We can now 
compute various $n$-point functions, by taking derivatives of 
$F = \sum_g F_g$ (where $F_g = F_g^{Kont}$ for $g\geq 1$)  
w.r.t. the $J_k$.
In doing so, we must keep in mind that $b$
depends implicitly on the $J_k$ through  the constraint \eq{constraint-3}. 
On the other hand, the model should be fixed, i.e. $a$ is fixed 
for given $N$ (this will be done below).
However, for derivatives up to second order we can as
well consider $u_0$ resp. $b$ as an independent variable as 
discussed above, 
taking advantage of \eq{constr-implicit-2}. 
This simplifies some of the computations below.

\subsection{Renormalization and finiteness}
\label{sec:renormaliz}

We can now determine the required renormalization of $a$, by
considering the one-point function. 
Using  \eq{F0tilde-2}, \eq{constr-implicit-2} and \eq{constraint-3}, 
the genus zero
contribution is
\bea
\langle \tilde \phi_{kk}\rangle_{g=0} &=&
\frac{i\la}{J_k}\,\frac{\dd{}}{\dd J_k } \,F_0 \nn\\
&=& \frac 1{i\la} \sqrt{J_k^2+2b}  
 + \frac 1{i\la}\frac {a-b}{\sqrt{J_k^2+2b}}
 - (i\la) \sum_{j=0}^N \frac{(J_k^2+2b)^{-1/2}}
{\sqrt{J_k^2+2b}+\sqrt{J_j^2+2b}} \nn\\
&& + \frac{\dd{ F_0}}{\dd b } \,\frac{i\la}{J_k}\,\frac{\dd{ b}}{\dd J_k }\nn\\
&=& \frac 1{i\la}\sqrt{J_k^2+2b}
 + (i\la) \sum_{j=0}^N\frac{1}{\sqrt{J_k^2+2b}\sqrt{J_j^2+2b}
   + (J_j^2+2b)}.
\label{F-derivative-1}
\eea
This is manifestly finite for the values \eq{J-explicit} of $J_j$ 
provided $b$ is finite, which
strongly suggests that a renormalization of the model is achieved
if $b$ is finite. To establish this, we have to check that 
the higher genus contributions are then also finite. 
Indeed,
\be
I_p = -(2p-1)!!\sum_{i=0}^{N}\frac 1{(\tilde m_i^2-2u_0)^{p+\frac 12}} 
= -(2p-1)!! \la^{2(2p+1)/3}
\sum_{i=0}^{N}\frac 1{(J_i^2 + 2b)^{p+\frac 12}} 
\label{I-p-sum}
\ee
is finite for $p \geq 1$.
We also obtain from \eq{constraint-3} that
\be
\frac{1}{J_k}\,\frac{\partial}{\partial J_k} b
=  -\frac {\la^2}{(J_k^2 + 2b)^{3/2}} 
- \sum_{i=0}^N \frac {\la^2}{(J_i^2 + 2b)^{3/2}} 
\frac{1}{J_k}\,\frac{\partial}{\partial J_k} b,
\ee
hence
\be
\frac{1}{J_k}\,\frac{\partial}{\partial J_k} b
= -\frac {\la^2}{(J_k^2 + 2b)^{3/2}}\, \frac {1}{1 - I_1}
\ee
is finite since  $1-I_1\neq 0$ for small $|i\la|$.
Therefore taking into account the implicit dependence of $b$ on
$J_k$, we see that
\be
\frac{1}{J_k}\,\frac{\partial}{\partial J_k} I_p 
=   \frac{(2p+1)!! \la^{2(2p+1)/3}}{(J_k^2 + 2b)^{p+\frac 12}}
(1+  \frac{1}{J_k}\,\frac{\partial}{\partial J_k} b)
\ee
is also finite.
Together with the structure of the higher genus contributions $F_g$ 
stated below \eq{F-higher} as found by \cite{Itzykson:1992ya}, this
implies that 
\begin{itemize}
\item[] {\em  All derivatives of $F_g$  w.r.t. 
$J_k$ for $g \geq 0$ as well as all $F_g$ for $g \geq 1$
are finite and have a well-defined limit $N \to \infty$, 
provided $b$ is finite.}
\end{itemize}
Since the connected $n$-point functions are given by the derivatives
of $F = \sum_{g \geq 0} F_g$ w.r.t. $J_k$, this implies that
all contributions in a genus expansion of the
correlation functions for diagonal entries 
$\langle \phi_{kk} ... \phi_{ll}\rangle$ are finite and well-defined. 
The general non-diagonal correlation functions are
discussed in section \ref{sec:general-correl}, 
and also turn out to be finite 
for arbitrary genus $g$ provided $b$ is finite.
Putting these results together we have
established renormalizability of the model
to all orders in a genus expansion, i.e.
\begin{itemize}
\item[]
{\em The (connected) genus $g$ contribution 
to any given $n$-point function is finite and has a 
well-defined limit $N \to \infty$ for all $g$, provided $b$ is finite.}
\end{itemize}   
Moreover, they can in principle 
be computed explicitly using the above formulas.
In particular, since any contribution to 
$F_g$  has order at least $\la^{4g-2}$, 
this implies  renormalizability of 
the perturbative expansion to any order in $\la$.

Next we show that the renormalization conditions\footnote{we 
will not bother 
to impose \eq{renorm-cond} exactly, 
since $\mu^2$ does not require renormalization} 
$\langle\phi_{00}\rangle = \langle\tilde
\phi_{00}\rangle-\frac{J_0}{i\la} =0$  \eq{renorm-cond} 
has indeed a solution
with finite $b$.  At genus 0, 
this amounts to
\bea
J_0 (\sqrt{1+2\frac{b}{J_0^2}}-1) 
&=& -(i\la)^2 \sum_{j=0}^{N} \frac{1}
{(J_0^2+2b)^{1/2}(J_j^2+2b)^{1/2} +(J_j^2+2b) } \nn\\
&=& -(i\la)^2 \Big(\sum_{j=0}^{N} \frac{1} {J_0 J_j +J_j^2 } - O(b)\Big)\nn\\
&\approx& 
- (i\la)^2 
 \Big(\frac{1}{4\pi}\int_{J_0}^\infty dJ\frac 1{J_0 J + J^2} - O(b)\Big) \nn\\
 &=& -(i\la)^2 \Big( \frac {\ln 2}{4\pi} \frac 1{J_0} - O(b)\Big) .
\label{renormaliz-cond-2}
\eea
approximating the sum by an integral.
The lhs is an unbounded increasing function of $b$ starting at 0, 
while the rhs is decreasing and positive for real $\la$. Therefore
\eq{renormaliz-cond-2} 
has a unique solution 
\be
b  = b(\la) > 0.
\ee
The higher-genus terms contribute only to higher order in $\la$, 
and therefore cannot change this conclusion
for small enough $|\la|$.
In particular, $b \to 0$ as $\la \to 0$.
This determines $b$ to be finite in the properly renormalized
model, which represents together with the mass and coupling constant
one of the free physical,
properly renormalized parameters of the model.
This in turn determines the counterterm $a$ through \eq{b-def}, 
\be
a = b +  \la^{4/3}I_0
\ee
which is therefore logarithmically divergent
in $N$, as is $I_0$. This will be worked out in more detail below.

\paragraph{Analytic continuation to real $(i\la)$.}

It is easy to see that $b(\la)$ is analytic
in $\la$ near the origin, and allows an analytic continuation 
to real $(i\la)$. The analytic property of $b(\la)$ can be understood by
considering the inverse function 
\bea
 (i\la)^2 &=& -J_0 (\sqrt{1+2\frac{b}{J_0^2}}-1)
\left(\sum_{j=0}^{N} \frac{1}
{(J_0^2+2b)^{1/2}(J_j^2+2b)^{1/2} 
+(J_j^2+2b)}\right)^{-1}  \nn\\
&\approx& - J_0^2 (\sqrt{1+2\frac{b}{J_0^2}}-1)
(\frac{4\pi}{\ln 2} + O(b)) \nn\\
&=& -\frac{4\pi}{\ln 2}\, b (1+O(b)).
\label{renormaliz-cond-3}
\eea
as shown above.
This means that 
\be
b = b(\la) = (i\la)^2 f((i\la)^2)
\ee
where $f$ is an analytic function in $(i\la)^2$ near the origin.
The higher genus contributions are of higher order in $(i\la)^2$ 
and do not change this conclusion. This implies that 
$b(\la)$ can be analytically continued to  
to real $\la' =(i\la)$ in some neighborhood of the origin.

This can be seen more explicity. 
Consider again the first line of \eq{renormaliz-cond-3}
for any real $\la' =(i\la)$. 
For $2b \in [-J_0^2,0]$, the function
$-J_0(\sqrt{1+2\frac{b}{J_0^2}}-1)$ covers the interval 
$[J_0,0]$, while the term $()^{-1}$ 
covers the interval\footnote{the lower bound is approximate} 
$[0,\frac{\ln 2}{4\pi J_0}]$. Therefore there exists a solution $b(\la')$
for real $\la' =(i\la)$ provided $|\la'|$ 
is small enough.

If  $J_0^2+2b < 0$,
some of the eigenvalues $m_i$ of the Kontsevich model become 
imaginary. It seems unlikely that there is a 
meaningful solution of \eq{renormaliz-cond-3} in that case. 
We will not pursue this any further in this paper.

\subsection{Small coupling expansion and checks}
\label{sec:expand}

We want to verify that the results derived from the Kontsevich 
integrals coincide with the leading non-trivial perturbative 
computations. Only genus 0 contributes here.

Consider first the one-point function 
$\langle \phi_{kk}\rangle$. Using the fact that $b = O(\la^2)$
as shown above, we have using \eq{F-derivative-1} 
\bea
\langle \phi_{kk}\rangle &=&
\frac{J_k}{i\la} (\sqrt{1+2\frac{b}{J_k^2}}-1)
 +(i\la) \sum_{j=0}^{N} \frac{1}
{(J_k^2+2b)^{1/2}(J_j^2+2b)^{1/2} +(J_j^2+2b) } \nn\\
&=& \frac{1}{i\la}\frac{b}{J_k}+(i\la) \sum_{j=0}^{N} \frac{1}
{J_k J_j +J_j^2} + O(\la^3) \nn\\
&=&  \frac{1}{J_k}\Big(\frac{b}{i\la}+(i\la)\sum_{j=0}^{N} 
 (\frac{1}{J_j} -\frac{1}{J_k +J_j})\Big) + O(\la^3).
\label{renormaliz-cond-4}
\eea
On the other hand, the constraint \eq{constraint-3} 
is to lowest order\footnote{note that all higher order corrections are finite}
\be
(b -a) =  \la^2\sum_{i=0}^N \frac 1{J_i} + O(\la^4)
\label{constraint-5}
\ee
and combined with the above this gives
\bea
\langle \phi_{kk}\rangle &=& 
\frac{1}{J_k}\Big(\frac{b}{i\la} + \sum_{j=0}^{N} 
 (\frac{(i\la)}{J_j} -\frac{(i\la)}{J_k +J_j})\Big) + O(\la^3)\nn\\
 &=& \frac{1}{J_k}\Big(\frac a{i\la} 
-\sum_{j=0}^{N} \frac{i\la}{J_k +J_j}\Big)\,\, + O(\la^3)\nn\\
&\approx& \frac {1}{J_k} \left(\frac a{i\la} - \frac {i\la}{4\pi}
  \ln\frac{J_k+J_N}{J_k+J_0}   \right) \,\, + O(\la^3).
\label{1point-1}
\eea
This agrees precisely with a perturbative computation, as
shown in section \ref{sec:perturbative}.
We note in particular that the dependence on $k$ of
$\langle \phi_{kk}\rangle$ is nontrivial, and we cannot require 
that $\langle \phi_{kk}\rangle =0$ for all $k$.
Setting $\langle \phi_{00}\rangle = 0$ determines
the required one-loop counterterm
\be
a = (i\la)^2 \sum_{j=0}^{N} \frac{1}{J_0 +J_j}\,\, + O(\la^4)
\approx \frac {(i\la)^2}{4\pi} \ln\frac{J_0+J_N}{2 J_0} \,\,+ O(\la^4)
\ee
which gives
\bea
b &=& (i\la)^2 \sum_{j=0}^{N} (\frac{1}{J_0 +J_j} 
  - \frac 1{J_j})\,\, \nn\\
&\approx& \frac {(i\la)^2}{4\pi}\Big(\ln\frac{J_0+J_N}{J_0+J_0} 
- \ln\frac{J_N}{J_0}\Big) \,\,+ O(\la^4) \nn\\
&\approx& -\frac {(i\la)^2}{4\pi}\ln 2 \,\,+ O(\la^4)
\label{b-perturb}
\eea
for large $N$.

\paragraph{The propagator $<\phi_{kl}\phi_{lk} >$.}

We can use \eq{2point-eom}
for the lowest-order correction to the 2-point function
$<\phi_{kl}\phi_{lk} >$ for $k \neq l$. First 
we observe that
\bea 
\frac{\langle \phi_{kk} - \phi_{ll}\rangle}{J_k^2 - J_l^2}
+ \frac{\langle \phi_{kk} + \phi_{ll}\rangle}{(J_k + J_l)^2}
&=& \frac 2{(J_k + J_l)^2(J_k-J_l)}
 \Big(J_k\langle \phi_{kk}\rangle - J_l\langle \phi_{ll}\rangle\Big) \nn\\
&=& \frac{2i\la}{(J_k + J_l)^2} \sum_{j=0}^{N} 
  \frac{1}{J_k +J_j}\frac{1}{J_l +J_j} \,\,+ O(\la^3)
\eea
using \eq{1point-1}.
Therefore  \eq{2point-eom} gives
\bea
\Big<\phi_{kl}\phi_{lk}\Big>
&=& \frac{2}{J_k + J_l} 
- 2(i\la) \frac{\langle \phi_{kk} + \phi_{ll}\rangle}{(J_k + J_l)^2}
+ \frac{4(i\la)^2}{(J_k + J_l)^2} \sum_{j=0}^{N} 
  \frac{1}{J_k +J_j}\frac{1}{J_l +J_j}
\,\,+ O(\la^4).  \nn\\
\label{2point-eom-2}
\eea
This again coincides with the diagrammatic result
\eq{2point-eom-2-pert}.

\paragraph{The propagator $\langle\phi_{ll}\phi_{kk}\rangle$.}

As a further example, consider the 2-point function 
$\langle\phi_{ll}\phi_{kk}\rangle$ for $k \neq l$, 
which vanishes in the free case.
To compute it from the effective action, we need in principle
\bea
\langle\tilde\phi_{ll}\tilde\phi_{kk}\rangle - \langle\tilde\phi_{kk}\rangle
\langle\tilde\phi_{ll}\rangle 
&=& \frac{i\la}{J_l}\frac{\partial}{\partial J_l}
\frac{i\la}{J_k}\frac{\partial}{\partial J_k}(F_0 + F_1 + ...) .
\eea
Even though this corresponds to a nonplanar diagram with external legs, 
it is  obtained by
taking derivatives of a closed genus 0 ring diagram.
Therefore we expect that only $F_0$ will contribute, 
and indeed the derivatives of $F_1$ contribute only to order $\la^{4}$.
We need 
\bea
&&\frac{i\la}{J_l}\frac{\partial}{\partial J_l}
\frac{i\la}{J_k}\frac{\partial}{\partial J_k} F_0 = \nn\\
&=& -\frac{i\la}{J_l}\frac{\partial}{\partial J_l}
 \frac{1}{\sqrt{J_k^2+2b}}
\left(\la^{-2}(J_k^2+2b) + \la^{-2} (a-b)
 + \sum_{j=0}^N \frac{1}{\sqrt{J_k^2+2b}
   +\sqrt{J_j^2+2b}} \right) \nn\\
&=& - \frac{(i\la)^2}{\sqrt{J_k^2+2b}}
{J_l}\frac{\partial}{\partial J_l}
 \left(\sum_{j=0}^N \frac{1}{\sqrt{J_k^2+2b}
   +\sqrt{J_j^2+2b}} \right) \nn\\
&=& (i\la)^2\frac{1}{\sqrt{J_k^2+2b}}\frac{1}{\sqrt{J_l^2+2b}}
 \left(\frac{1}{\sqrt{J_k^2+2b} +\sqrt{J_l^2+2b}}\right)^2  
\eea
Therefore\footnote{This computation was again simplified by
ignoring  the implicit dependence of  
$b$ on the $J_i$ taking advantage of \eq{constr-implicit}. 
This is no longer possible for higher 
derivatives.}
 to lowest order we obtain
\bea
\langle\phi_{ll}\phi_{kk}\rangle 
= \langle\phi_{kk}\rangle \langle\phi_{ll}\rangle 
 + \frac{(i\la)^2}{J_k\, J_l}\left(\frac{1}{J_k + J_l}\right)^2,
\label{phil-phik-kont}
\eea
in complete agreement with the perturbative computation 
\eq{2point-nonplanar}.

\subsection{Approximation formulas for finite coupling}
\label{sec:asymptotic}

In this section we derive some closed formulas which 
are appropriate for finite coupling $\la$, in the large 
$N$ limit.
This is done by approximating the various sums by integrals.
Using \eq{I-p-sum}, we have 
\bea
I_0  &=& - \la^{2/3}
\sum_{i=0}^{N}\frac 1{(J_i^2 + 2b)^{\frac 12}} \,\,
\approx -\frac{\la^{2/3}}{\sqrt{2b}}
\frac{\sqrt{2b}}{4\pi} 
\int_{x_0}^{x_N} dx \frac 1{(x^2 + 1)^{1/2}}  \nn\\
&=& -\frac{\la^{2/3}}{4\pi} \ln(\frac{x_N+\sqrt{1+x_N^2}}{x_0+\sqrt{1+x_0^2}})
\eea
where 
\be
x_n = \frac{4\pi}{\sqrt{2b}}\, (n+ \frac{1+\mu^2}2), \quad
dx = \frac{4\pi}{\sqrt{2b}}\, dn \, .
\label{xn-def}
\ee
Furthermore,  we will need
\bea
I_1 &=&  - \la^{2}
\sum_{i=0}^{N}\frac 1{(J_i^2 + 2b)^{\frac 32}} \nn\\
&\approx& -\frac{\la^{2}}{(2b)^{3/2}} 
\frac{\sqrt{2b}}{4\pi} 
\int_{x_0}^{x_N} dx \frac 1{(x^2 + 1)^{3/2}} \nn\\
&=&  -\frac{\la^{2}}{2b} \frac{1}{4\pi} 
\Big(\frac{1}{\sqrt{1+x_N^{-2}}}-\frac{1}{\sqrt{1+x_0^{-2}}}\Big)\nn\\
&\approx& -\frac{\la^{2}}{2b} \frac{1}{4\pi} 
\Big(1-\frac{1}{\sqrt{1+x_0^{-2}}}\Big).
\label{I1-asymptot}
\eea
This is valid also for  $b <0$ (by analytic continuation)
as long as $1+x_0^{-2} >0$.
Similarly, all $I_p$ can be approximated by 
elementary, convergent integrals.

Consider again equation \eq{renormaliz-cond-2}, which determines 
$b$ as a function of
the coupling constant at genus 0. 
It can be written for finite $b$ using the above approximation as
\bea
J_0 (\sqrt{1+2\frac{b}{J_0^2}}-1) 
&=& -(i\la)^2 \sum_{j=0}^{N} \frac{1}
{(J_0^2+2b)^{1/2}(J_j^2+2b)^{1/2} +(J_j^2+2b) } \nn\\
&\approx& - \frac{(i\la)^2}{2b} 
\frac{\sqrt{2b}}{4\pi} 
\int_{x_0}^{x_N} dx \frac{1}
{\sqrt{x_0^2+1}\sqrt{x^2+1} +x^2+1 } \nn\\
&\approx& - \frac{(i\la)^2}{4\pi} \frac 1{\sqrt{2b}} 
\frac 1{x_0}\ln(1+\frac{1}{\sqrt{1+x_0^{-2}}})
\label{renormaliz-cond-7}
\eea
using
\bea
\int_{x_0}^{\infty} dx 
\frac{x_0}{\sqrt{x_0^2+1}\sqrt{x^2+1} +(x^2+1)}
&=& \ln(1+\frac{1}{\sqrt{1+x_0^{-2}}}).
\eea
Therefore \eq{renormaliz-cond-7} becomes 
\bea
J_0^2\Big(\sqrt{1+\frac{2b}{J_0^2}}-1\Big) 
&=&- \frac{(i\la)^2}{4\pi} 
\ln\Big(1+\frac{1}{\sqrt{1+\frac{2b}{J_0^2}}}\Big) . 
\label{renormaliz-cond-5}
\eea
This implies again that 
there exists a solution $b = b(\la)$ not only for real $\la$, but also for 
real $(i\la)$ at least in some neighborhood of the origin.
We can also recover the leading perturbative result \eq{b-perturb}, 
\be
2b=-\frac{(i\la)^2  }{4\pi}\frac 1{x_0^2}
\frac 1{\sqrt{1+x_0^{-2}}-1}
\ln(1+\frac{1}{\sqrt{1+x_0^{-2}}})
= -\frac{(i\la)^2  }{4\pi} 2\ln 2 + O(x_0^{-2}).
\label{renormaliz-cond-6}
\ee
Using these and similar
expressions, one can obtain explicit formulas
for the genus expansion of $F$
and the correlators considered in the previous section,
which are appropriate for finite coupling. Since there is no 
particular difficulty we will not write them down explicitly.

\subsection{Critical line and instability.}
\label{sec:critical}

We have seen that for small enough coupling $|\la|$, 
the free energy $F = F_0 + F_1 + ...$
is regular and  finite for any given genus in the renormalized
model (i.e. for finite $b$), since all $I_k$ with $k \geq 1$
are finite {\em provided} $I_1 \neq 1$. 

However, as is manifest in the explicit formulas for $F_g$ at higher
genus \eq{F-0-IZ-1} ff., there is a singularity at $I_1=1$.
Using \eq{I1-asymptot}, 
this critical point is given by
\bea
1 &=& I_1 
= \frac{(i\la)^{2}}{2b} \frac{1}{4\pi} 
\Big(1-\frac{1}{\sqrt{1+x_0^{-2}}}\Big) \nn\\
\frac{8\pi b}{(i\la)^{2}}  &=& 
\Big(1-\frac{1}{\sqrt{1+x_0^{-2}}}\Big) 
= \frac 12 x_0^{-2} + O(x_0^{-4})
\label{critical-1}
\eea
for large $N$. The lhs is negative for both real $\la$ 
and real $i\la$, as can be seen e.g. from \eq{renormaliz-cond-3}.
Since $x_0^{-2} = \frac{2b}{4\pi^2(1+\mu^2)^2}$ 
\eq{xn-def}, \eq{critical-1} has a solution 
only for negative $b$ i.e. real $(i\la)$.
Combining this with \eq{renormaliz-cond-5}
which is valid for finite $b$, we get
\bea
\Big(1-\frac{1}{\sqrt{1+x_0^{-2}}}\Big) 
= -\frac{\ln(1+\frac{1}{\sqrt{1+x_0^{-2}}})}
{x_0^2(\sqrt{1+x_0^{-2}}-1)} 
\eea
which has a unique solution $x_0^{-2} \approx -0.873759$.
Inserting this into \eq{critical-1} gives
\bea
\frac{4\pi^2(1+\mu^2)^2}{2b} \frac{8\pi b}{(i\la)^{2}}    
&=& x_0^2\frac{8\pi b}{(i\la)^{2}} 
= x_0^2 \,\Big(1-\frac{1}{\sqrt{1+x_0^{-2}}}\Big) \approx 2.07665 \nn\\
\frac{1+\mu^2}{i\la} &\approx& \pm 0.0646989
\label{criticalline}
\eea
This\footnote{Recall that this is obtained
imposing the renormalization conditions $\langle \phi_{00}\rangle =0$ 
at genus 0. }
 indicates that for the $\phi^3$ model with real coupling 
constant $\la' = i\la$ stronger than this critical coupling, 
the model becomes unstable.
This is very reasonable, since the potential is unbounded, and the 
potential barrier around the local minimum becomes weaker 
for stronger coupling. Therefore this critical line
could be interpreted as 
the point where the quantum fluctuations of $\phi$ 
become large enough to see the global instability, so that the 
field ``spills over'' the potential barrier. 
Similar transitions 
for a cubic potential are known e.g. for
the ordinary matrix models, but may also be relevant 
in the context of string field theory
and tachyon condensation \cite{Sen:1999nx,Gaiotto:2003yb}.
In particular, it is interesting to note that this singularity
occurs simultaneously for each given genus, which suggest
that some double-scaling limit near this 
critical point can be taken, again in analogy with the 
usual matrix models (for a review, see
e.g. \cite{DiFrancesco:1993nw}). Again, such a 
scaling limit for the 
Kontsevich model is discussed in \cite{Itzykson:1992ya}. We leave this 
issue for future work.

\subsection{General $n$-point functions}
\label{sec:general-correl}

Finally we show that all contributions in the genus expansion (and therefore
perturbative expansion) of the expectation values of any $n$-point
functions of the form
\be
\langle \phi_{i_1 j_1} ....  \phi_{i_n j_n}\rangle 
\label{correl}
\ee
have a well-defined and finite limit as $N \to \infty$
provided $b$ is finite, which means
that the model is fully renormalized.

In view of \eq{action-kontsevich-lin},
the insertion of a factor $\tilde\phi_{ij}$ can be obtained by acting
with the derivative operator $2(i\la)
\frac{\partial}{\partial J^2_{ij}}$ on $Z(J)\,$ resp. 
$F_g(J)$ for fixed genus $g$. We use $J^2$ rather than $J$ to
simplify the notation, which is not a problem
defining the inverse by the local diffeomorphism
given by the positive square-root of a Hermitian matrix with distinct 
positive eigenvalues (alternatively one could
act with  $(-i\la\frac{\partial}{\partial a_{ij}} + \frac 1{i\la}
J_{ji})$ promoting $a$ to
a matrix, see \eq{action-kontsevich}).
Since $Z(J)\,$ and $F_g(J)$ depend only on the eigenvalues of $J$, 
we can diagonalize $J$ as $J = U^{-1} \diag(J_a) U$, and
rewrite these derivatives as 
\be
\frac{\partial}{\partial J^2_{ij}} 
= \sum_a\frac{\partial J_a}{\partial J^2_{ij}}
\frac{\partial}{\partial J_{a}} 
+  \sum_{\a}\frac{\partial V_{\a}}{\partial J^2_{ij}}
\frac{\partial}{\partial V_{\a}}
\label{derivat-general}
\ee
where $V_{\a}$ denotes suitable coordinates on the coset space 
$U(N)/(\prod U(1))$ near the origin. 
Now for any given correlation function of type \eq{correl}, 
let $k$ denote the highest index involved. Then taking 
derivatives of the type \eq{derivat-general} amounts to considering 
matrices $J$ of the form
\be
J = \left(\begin{array}{ccc} J_{k\times k} &\vline & 0 \\
                   \hline 
                      0   & \vline & \diag(J_{k+1},... J_N)\end{array}\right)
\ee
We can assume that $J_{k\times k}$ is a small perturbation around
$\diag(J_{1},... J_k)$ with distinct increasing eigenvalues. 
Therefore only the first $k$ eigenvalues of $J$ are deformed, while 
the higher eigenvalues $J_{k+1},... J_N$ are fixed and given by 
\eq{J-explicit}.
This means that the diagonalization and the
transformation \eq{derivat-general}
takes place only in the upper-left $k\times k$ block, and 
is independent of $N$ for $N > k$.  In particular, only the coset
coordinates $V_{\a}$ for $U(k)/(\prod U(1))$ enter.  Therefore
acting with multiple operators of type \eq{derivat-general} on
$F_g(J)$ (resp. $Z(J)$)
produces a result which is finite and independent of $N$ except
possibly through the terms
$\frac{\partial}{\partial J_{a}} ... \frac{\partial}{\partial J_{b}}
F_g(J)$ (resp. 
$\frac{\partial}{\partial J_{a}} ... \frac{\partial}{\partial
  J_{b}}Z(J)$), 
which in turn were shown to be
finite and convergent  for any $g$ in section \ref{sec:renormaliz}. 
This completes the proof that each genus $g$ contribution to 
the general (connected) 
correlators $\langle \phi_{i_1 j_1} ....  \phi_{i_n j_n}\rangle $
is finite and convergent as $N \to \infty$.
This implies in particular (but is stronger than) renormalizability of 
the perturbative expansion to any order in $\la$.

\section{Perturbative computations}
\label{sec:perturbative}

\begin{figure}[htpb]
\begin{center}
\epsfxsize=2in
   \epsfbox{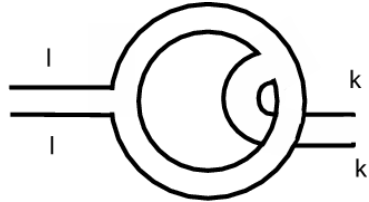}
\end{center}
 \caption{one-loop contribution to $\langle\phi_{ll}\phi_{kk}\rangle $}
\label{fig:nonplanar}
\end{figure}

We write the action \eq{action-tilde} as
\bea
\tilde S &=& Tr \Big(\frac 14 (J \phi^2 + \phi^2 J)
+ \frac{i\la}{3!}\;\phi^3 - \frac{a}{i\la} \phi \Big) \nn\\
&=&  Tr( \frac 12 \phi^i_j \; G^{j;l}_{i;k}\;  \phi^k_l 
+ \frac{i\la}{3!}\;\phi^3 - \frac{a}{i\la} \phi)
\eea
where the kinetic term is
$ G^{j;l}_{i;k}  = \frac 12 \delta^i_l \delta^k_j (J_i+J_j)$, 
and the propagator is
\be
\Delta^{i; k}_{j;l} = \langle \phi^i_j \phi^k_l\rangle 
 = \delta^i_l \delta^k_j \frac 2{J_i+J_j}
= \delta^i_l \delta^k_j \frac {1/(2\pi)}{i + j + (\mu^2\theta+1)}.
\label{propagator}
\ee
In 2 dimensions, the only divergent primitive graph is 
the tadpole, which requires the counterterm $a$. 
A one-loop computation gives
\bea
\langle \phi_{ii} \rangle 
&=&\frac{a}{i\la} \frac 1{J_i}  - \frac{i\la}2\,  
\frac 1{J_i} \sum_{k=0}^N  \frac 2{J_i+J_k}  
\approx -\frac {i\la}{J_i} \left(\frac a{\la^2} + \frac {1}{4\pi}
  \ln\frac{J_i+J_N}{J_i+J_0}   \right)\nn\\
&=& -\frac 1{2\pi}\frac {i\la}{2i+\mu^2+1} \left(\frac a{\la^2} + \frac {1}{4\pi}
  \ln\frac{N+i+\mu^2+1}{i+\mu^2+1} \right)
\label{onepoint-oneloop}
\eea
In particular, $\langle \phi_{00}\rangle =0$ implies
\be
a + \frac {\la^2}{4\pi} \ln\frac{N+\mu^2+1}{\mu^2+1} =0.
\label{a-pert-1}
\ee
Next we compute the leading contribution to the 
2-point function  $\langle\phi_{ll}\phi_{kk}\rangle$ for $l\neq k$,
which vanishes at tree level. The leading contribution 
comes from the nonplanar graph in figure \ref{fig:nonplanar},
which gives
\bea
\langle\phi_{ll}\phi_{kk}\rangle 
= \langle\phi_{kk}\rangle \langle\phi_{ll}\rangle 
+ \frac 14 \frac{(i\la)^2}{J_k\, J_l}\left(\frac{2}{J_k + J_l}\right)^2  
\label{2point-nonplanar}
\eea
(for $l\neq k$) indicating the symmetry factors, 
where the disconnected contributions 
are given by \eq{onepoint-oneloop}.
This is in complete agreement with the result 
\eq{phil-phik-kont} obtained from the 
Kontsevich model approach.

\begin{figure}[htpb]
\begin{center}
\epsfxsize=3.5in
  \vspace{0.2in} 
   \epsfbox{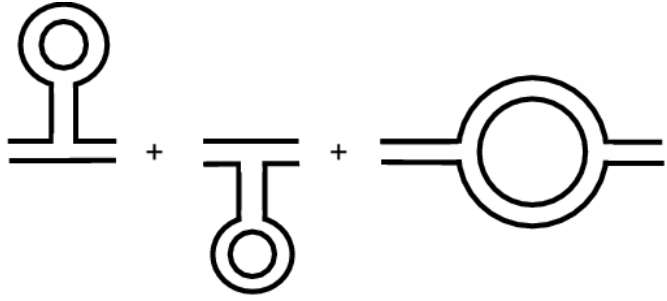}
\end{center}
 \caption{one-loop contribution to $\langle\phi_{kl}\phi_{lk}\rangle $}
\label{fig:planarprop}
\end{figure}

Similarly, the leading contribution to the 
2-point function  $\langle\phi_{kl}\phi_{lk}\rangle$ for $l\neq k$,
has the contribution indicated in figure \ref{fig:planarprop}, 
which give the result
\bea
\Big<\phi_{kl}\phi_{lk}\Big>
&=& \frac{2}{J_k + J_l} 
- 2(i\la) \frac{\langle \phi_{kk} + \phi_{ll}\rangle}{(J_k + J_l)^2}
+ \frac{4(i\la)^2}{(J_k + J_l)^2} \sum_{j=0}^{N} 
  \frac{1}{J_k +J_j}\frac{1}{J_l +J_j}
\,\,+ O(\la^4).  \nn\\
\label{2point-eom-2-pert}
\eea
The first term 
is the free propagator, the second term the tadpole contributions
including counterterms, and
the last them the one-loop contribution in figure \ref{fig:planarprop}.
This is in complete agreement with the result 
\eq{2point-eom-2} obtained from the 
Kontsevich model approach.

\begin{figure}[htpb]
\begin{center}
\epsfxsize=1.8in
  \vspace{0.2in} 
   \epsfbox{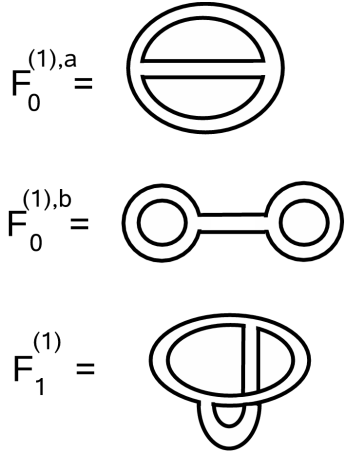}
\end{center}
 \caption{leading contribution to $F_0, \, F_1$}
\label{fig:Fplanar}
\end{figure}

One can also check the leading terms for $F_0,F_1$ 
in \eq{F0-explicit}, \eq{F1-explicit} explicitly.
The lowest order contribution to
$F_0$ is given by the planar
diagrams in figure \ref{fig:Fplanar}, 
which are given by 
\bea
F_0^{(1),a} &=& (i\la)^2 \sum_{i,j,k}\frac 2{J_i+J_j}\, \frac 2{J_j+J_k}\, 
\frac 2{J_k+J_i}, \nn\\
F_0^{(1),b} &=& (i\la)^2 \sum_{i,j,k}\frac 2{J_i+J_j} \,\frac 2{2J_i} \,\frac 2{J_i+J_k}.
\eea
It is not difficult to see that 
\be 
F_0^{(1)} := \frac{(i\la)^2}2\,
\langle \frac 16 Tr \phi^3 \frac 16 Tr \phi^3 \rangle_0
 = \frac 1{72}(3 F_0^{(1),a} + 9 F_0^{(1),b}) 
=  \frac{(i\la)^2}6  \left(\sum_i\frac 1{J_i}\right)^3
= \frac {\la^2}6\, t_0^3
\ee
(setting $a=0$ to this lowest order),
which indeed coincides with the first term in \eq{F0-explicit}.
Similarly,
the lowest order contribution to
$F_1$ is given by the nonplanar diagrams in figure 
\ref{fig:Fplanar}, which sum up to
\bea
F_1^{(1)} := \frac{(i\la)^2}2\,
\langle \frac 16 Tr \phi^3 \frac 16 Tr \phi^3 \rangle_1
 = \frac{(i\la)^2}{24} \sum_i \left(\frac 1{J_i}\right)^3 
= \frac {\la^2}{24}\,t_1
\eea
again in agreement with the first term in \eq{F1-explicit}.
This illustrates the remarkable role 
of the variables $t_r$ \eq{var-defs}
in the Kontsevich model.

We also would like to point out that the perturbative expressions obtained
in this model are very similar to those considered by Connes and
Kreimer in
\cite{Connes:1998qv}. It might be interesting to study
their Hopf-algebraic formulation of renormalization 
in this nonperturbative framework.

\section{Discussion and conclusion}

In this paper, we have shown that the selfdual NC
$\phi^3$ model can be mapped to the Kontsevich (matrix) model,
for a suitable external source resp. eigenvalues of the latter.
This map works for any even dimensions. 
We concentrate on the 2dimensional case in this paper, leaving the 
case of 4 dimensions for a future publication.
We showed how to use known results for the 
Kontsevich model to quantize and renormalize 
the noncommutative $\phi^3$ model. This provides closed
expressions for each given genus $g$ in a genus expansion,
which are valid for finite nonzero coupling. 
The appropriate renormalization is found, and 
we have shown that the resulting contributions for each genus
are finite and well-defined 
for nonzero coupling. This implies but is stronger than
renormalization order-by-order in perturbation theory.
An instability is found 
if the real coupling constant reaches a critical
coupling, as expected for the $\phi^3$ model.

The techniques used in this paper are very powerful, but
more-or-less restricted to the $\phi^3$ interaction 
(note however the possible generalizations of the Kontsevich model
pointed out in \cite{Itzykson:1992ya,Kontsevich:1992,Kharchev:1991cu}). 
However, one important 
message is the fact that the required renormalization 
is determined by the genus 0 contribution only.  
This can be expected to hold more generally 
for scalar NC models. Since the genus 0 contribution should
accessible more easily and does include essential nonperturbative
information, this supports the strategy of applying  matrix-methods such as
those in \cite{Steinacker:2005wj,Steinacker:2005tf}
more generally in the NC case.

Perhaps the main gap in our treatment is the lack of control
over the {\em sum } over all genera $g$. While the contributions
for each genus are manifestly analytic in the coupling constant
$\la$, we have not shown that the sum over $g$ 
converges in a suitable sense. However, it seems very plausible that 
this is the case, and the sum defines an analytic function 
in $\la$ near the origin. This would amount 
to a full construction of the model.
It should  be possible to establish this using 
the relation with the KdV hierarchy or the relation with topological 
gravity, which is 
however beyond the scope of this paper. 

In view of the well-known relation  between NC field theory and string theory
in certain backgrounds \cite{Seiberg:1999vs}, 
our results are quite relevant also in that context.
Renewed interest arises through the recent discussions of  
the relevance of the Konsevich model e.g.
for open string field theory and open/closed string 
duality \cite{Gaiotto:2003yb},
and for branes on Calabi-Yau spaces \cite{Aganagic:2003qj}.

\paragraph{Acknowledgements}

We are grateful for useful discussions with R. Wulkenhaar and 
E. Langmann, and to J. Teschner for pointing out reference 
\cite{Gaiotto:2003yb}.
This work was supported  by the FWF project P16779-N02.

\bibliographystyle{diss}

\bibliography{mainbib}

\end{document}